\documentclass[useAMS,usenatbib]{mn2e}

\usepackage{graphicx}

\newcommand{\vsini}{$v \sin i$}

\newcommand{\geneva}{{\sc geneva}}

\newcommand{\logg}{$\log(g)$}

\newcommand{\logteff}{$\log(T_{\mathrm{eff}})$}
\newcommand{\logl}{$\log(L/L_{\odot})$}

\title[Discovery of magnetic fields in SPBs]{Discovery of magnetic fields in the
$\beta$\,Cephei star $\xi^1$\,CMa and in several Slowly Pulsating B
stars\thanks{Based on observations obtained at the European Southern
Observatory, Paranal, Chile (ESO programmes 072.D-0377(A), 073.D-0464(A),
073.D-0466(A), and 075.D-0295(A)).  }} \author[S. Hubrig et
al.]{S. Hubrig$^{1}$\thanks{E-mail: shubrig@eso.org},
  M. Briquet$^{2}$\thanks{Postdoctoral Fellow of the Fund for Scientific
    Research of Flanders (FWO)},
M. Sch\"oller$^{1}$, P. De Cat$^{3}$, G. Mathys$^{1}$, and C. Aerts$^{2}$ \\
$^{1}$European Southern Observatory, Casilla 19001, Santiago, Chile\\
$^{2}$Instituut voor Sterrenkunde, Katholieke Universiteit Leuven,
Celestijnenlaan 200B, B-3001 Leuven, Belgium\\ $^{3}$Koninklijke Sterrenwacht
van Belgi\"e, Ringlaan 3, B-1180 Brussel, Belgium }

\begin{document}

\date{Accepted 2006 Enero 99. Received 2006 Enero 98}

\pagerange{\pageref{firstpage}--\pageref{lastpage}} \pubyear{2006}

\maketitle

\label{firstpage}

\begin{abstract}
We present the results of a magnetic survey of a sample of eight $\beta$\,Cephei
stars and 26 Slowly Pulsating B stars with FORS\,1 at the VLT.  A weak
mean longitudinal magnetic field of the order of a few hundred Gauss is detected
in the $\beta$\,Cephei star $\xi^1$\,CMa and in 13 SPB stars.  The star
$\xi^1$\,CMa becomes the third magnetic star among the $\beta$\,Cephei stars.
Before our study, the star $\zeta$\,Cas was the only known magnetic SPB star.
All magnetic SPB stars for which we gathered several magnetic field measurements
show a field that varies in time.  We do not find a relation between the
evolution of the magnetic field with stellar age in our small sample.  Our
observations imply that $\beta$\,Cephei stars and SPBs can no longer be
considered as classes of non-magnetic pulsators, but the effect of the fields
on the oscillation properties remains to be studied.
\end{abstract}


\section{Introduction}
A long-term monitoring project aimed at asteroseismology of a large sample of
slowly pulsating B (SPB) stars and $\beta$\,Cephei stars was started
by researchers of the Institute of Astronomy of the University of Leuven several
years ago.  This initiative started after a number of new pulsating B stars had
been discovered from the Hipparcos mission (e.g.,
\citealt{DeCat2002,Briquet2003}).  $\beta$\,Cephei stars are short period
(3--8\,h) pulsating variables of spectral type O9 to B3 (8--20\,M$_\odot$) along
the main sequence with low-order pressure (p) and/or gravity
(g) modes.  SPB stars are situated in the H-\linebreak[0]R diagram just below
the $\beta$\,Cephei stars. They are less massive (3--9\,M$_\odot$) B-type stars
which show variability with periods of the order of one day due to multiperiodic
high-order low-degree g mode oscillations.  Thanks to the Hipparcos mission the
number of SPB stars increased from 12 to more than 80 \citep{Waelkens1998}.

The study of the pulsation properties of SPB and $\beta$\,Cephei stars requires
observations with a long time-base. Indeed, modes with closely spaced periods
occur.  As was pointed out by \citet{DeCat2002}, it is difficult to interpret
and explore these multiplets in view of the unknown internal rotation law
$\Omega(r)$ and given the dense theoretically predicted frequency spectra (e.g.,
\citealt{Aerts2006}).  Recently, \citet{Hasan2005} suggested that high-order
g modes are a probe of the internal magnetic field.
Their calculations show that frequency splittings of the order of 1\% may be due
to a poloidal field with a strength of order 100\,kG, buried in the deep interior
of the star.

\citet{Neiner2003a} made the first discovery of a magnetic field in an SPB
star. It concerned $\zeta$\,Cas whose dominant pulsation mode has a 
period P=1.56\,d.  The two other detections of a magnetic field in pulsating
B stars have been done for $\beta$\,Cephei stars: $\beta$\,Cep
\citep{Henrichs2000} and V2052\,Oph \citep{Neiner2003b}. The measured mean
longitudinal magnetic fields in all three stars are weak, less than
100\,G.  These discoveries motivated us to perform a systematic search for
magnetic fields in SPB and $\beta$\,Cephei stars, which started in 2003
with FORS\,1 at the VLT.  Our recent measurements in the hydrogen lines of
various types of stars of intermediate mass demonstrated that magnetic fields
can be measured with an error bar as small as 20\,G
\citep{Hubrig2006a,Hubrig2006b}, enabling us to diagnose even very weak mean
longitudinal fields.  While the last spectropolarimetric data for our program
have been released and reduced only in the last months, preliminary
results of the survey were already presented in 
\citet{Hubrig2005,Hubrig2006b}.  In this paper we present the whole 
sample and discuss the results of more than 80 magnetic field measurements.
\begin{table*}
\caption{
Fundamental parameters for the objects in our sample.
In the first two columns we give the HD number and another identifier.
In the following six columns we list spectral type, 
effective temperature, surface gravity, mass,
radius, and luminosity.
The final two columns give the fraction of the main sequence lifetime
and \vsini{}.
Uncertain parameter values are given in {\it italics}.
}
\label{table1}
\tabcolsep=5pt
\begin{center}
\begin{tabular}{rllrrrrcrr}
\hline
\multicolumn{1}{c}{HD} &
\multicolumn{1}{c}{Other} &
\multicolumn{1}{c}{Spectral} &
\multicolumn{1}{c}{\logteff{}} &
\multicolumn{1}{c}{\logg{}} &
\multicolumn{1}{c}{$M/M_\odot$} &
\multicolumn{1}{c}{$R/R_\odot$} &
\multicolumn{1}{c}{\logl{}} &
\multicolumn{1}{c}{$f$} &
\multicolumn{1}{c}{\vsini{}} \\
\multicolumn{1}{c}{} &
\multicolumn{1}{c}{Identifier} &
\multicolumn{1}{c}{Type} &
\multicolumn{1}{c}{} &
\multicolumn{1}{c}{} &
\multicolumn{1}{c}{} &
\multicolumn{1}{c}{} &
\multicolumn{1}{c}{} &
\multicolumn{1}{c}{} &
\multicolumn{1}{c}{[km/s]} \\
\hline
\multicolumn{10}{c}{$\beta$ Cephei stars} \\
\hline
 29248 & $\nu$ Eri & B2 III & 4.363$\pm$0.020 & 3.92$\pm$0.20 & 10.1$\pm$2.2 & 5.9$\pm$2.0 & 3.9$\pm$0.3 & 0.77$\pm$0.16 & 21$\pm$12  \\
 44743 & $\beta$ CMa & B1 II-III & {\it 4.421$\pm$0.020} & {\it 3.79$\pm$0.20} & {\it 13.5$\pm$1.5} & {\it 7.5$\pm$2.7} & {\it 4.4$\pm$0.2} & {\it 0.82$\pm$0.17} & 11$\pm$~\,7 \\
 46328 & $\xi^1$ CMa & B1 III & {\it 4.433$\pm$0.020} & {\it 3.83$\pm$0.20} & {\it 13.7$\pm$1.3} & {\it 7.1$\pm$2.4} & {\it 4.4$\pm$0.2} & {\it 0.76$\pm$0.18} & 20$\pm$~\,7 \\
111123 & $\beta$ Cru & B0.5 III, SB1 & 4.436$\pm$0.020 & 3.73$\pm$0.20 & {\it 14.1$\pm$0.9} & {\it 7.5$\pm$1.8} & {\it 4.4$\pm$0.1} & {\it 0.81$\pm$0.13} & 16$\pm$~\,9 \\
129929 & V836 Cen & B2 & 4.379$\pm$0.020 & 4.03$\pm$0.20 & 10.0$\pm$2.1 & 5.1$\pm$1.8 & 3.9$\pm$0.3 & 0.58$\pm$0.26 & 8$\pm$~\,5 \\
157056 & $\theta$ Oph & B2 IV, SB1 & 4.346$\pm$0.020 & 3.91$\pm$0.20 & 9.4$\pm$2.1 & 5.8$\pm$2.0 & 3.9$\pm$0.3 & 0.79$\pm$0.16 & 16$\pm$~\,8 \\
160578 & $\kappa$ Sco & B1.5 III, SB2 & 4.372$\pm$0.020 & 3.70$\pm$0.20 & 11.6$\pm$2.4 & 7.6$\pm$2.2 & 4.2$\pm$0.3 & 0.91$\pm$0.09 & 97$\pm$~\,1 \\
163472 & V2052 Oph & B2 IV-V & 4.351$\pm$0.020 & 3.86$\pm$0.20 & 10.0$\pm$2.2 & 6.4$\pm$2.2 & 4.0$\pm$0.3 & 0.84$\pm$0.13 & 63$\pm$~\,2 \\
\hline
\multicolumn{10}{c}{Slowly Pulsating B stars} \\
\hline
  3379 & 53 Psc & B2.5I V & 4.238$\pm$0.020 & 4.16$\pm$0.20 & 5.4$\pm$0.9 & 3.3$\pm$1.0 & 2.9$\pm$0.3 & 0.45$\pm$0.33 & 33$\pm$17  \\
 24587 & 33 Eri & B5 V, SB1 & 4.142$\pm$0.020 & 4.26$\pm$0.20 & 3.7$\pm$0.5 & 2.5$\pm$0.6 & 2.3$\pm$0.3 & 0.36$\pm$0.29 & 28$\pm$~\,1 \\
 26326 & GU Eri & B5 IV & 4.183$\pm$0.020 & 4.14$\pm$0.20 & 4.5$\pm$0.7 & 3.1$\pm$1.0 & 2.6$\pm$0.3 & 0.49$\pm$0.32 & 11$\pm$~\,6 \\
 28114 & V1143 Tau & B6 IV & 4.164$\pm$0.020 & 4.00$\pm$0.20 & 4.5$\pm$0.8 & 3.6$\pm$1.2 & 2.7$\pm$0.3 & 0.71$\pm$0.23 & 9$\pm$~\,5 \\
 34798 & YZ Lep & B5 IV-V, SB ? & 4.193$\pm$0.020 & 4.25$\pm$0.20 & 4.5$\pm$0.6 & 2.8$\pm$0.7 & 2.6$\pm$0.3 & 0.37$\pm$0.28 & 34$\pm$~\,2 \\
 45284 & BD-07 1424 & B8, SB2 & 4.167$\pm$0.020 & 4.40$\pm$0.20 & {\it 3.9$\pm$0.4} & {\it 2.4$\pm$0.3} & {\it 2.4$\pm$0.2} & {\it 0.19$\pm$0.16} & 71$\pm$~\,6 \\
 46005 & V727 Mon & B8 V & {\it 4.326$\pm$0.020} & {\it 4.43$\pm$0.20} & {\it 7.0$\pm$0.7} & {\it 3.2$\pm$0.3} & {\it 3.3$\pm$0.1} & {\it 0.10$\pm$0.07} & --\,~~~ \\
 53921 & V450 Car & B9 IV, SB2 & 4.137$\pm$0.020 & 4.23$\pm$0.20 & 3.7$\pm$0.5 & 2.6$\pm$0.7 & 2.3$\pm$0.3 & 0.39$\pm$0.30 & 17$\pm$10  \\
 74195 & $o$ Vel & B3 IV & 4.209$\pm$0.020 & 3.91$\pm$0.20 & 5.5$\pm$1.0 & 4.3$\pm$1.4 & 3.1$\pm$0.3 & 0.82$\pm$0.18 & 9$\pm$~\,5 \\
 74560 & HY Vel & B3 IV, SB1 & 4.210$\pm$0.020 & 4.15$\pm$0.20 & 4.9$\pm$0.8 & 3.1$\pm$1.0 & 2.8$\pm$0.3 & 0.46$\pm$0.32 & 13$\pm$~\,7 \\
 85953 & V335 Vel & B2 III & 4.266$\pm$0.020 & 3.91$\pm$0.20 & 6.8$\pm$1.4 & 4.9$\pm$1.7 & 3.4$\pm$0.3 & 0.81$\pm$0.18 & 18$\pm$10  \\
 92287 & V514 Car & B3 IV, SB1 & 4.216$\pm$0.020 & 4.00$\pm$0.20 & 5.4$\pm$1.0 & 3.9$\pm$1.3 & 3.0$\pm$0.3 & 0.70$\pm$0.23 & 41$\pm$16  \\
123515 & V869 Cen & B9 IV, SB2 & 4.079$\pm$0.020 & 4.27$\pm$0.20 & 3.0$\pm$0.4 & 2.2$\pm$0.5 & 2.0$\pm$0.3 & 0.36$\pm$0.28 & 6$\pm$~\,3 \\
131058 & $\zeta$ Cir & B3 V & 4.225$\pm$0.020 & 4.03$\pm$0.20 & 5.5$\pm$1.0 & 3.8$\pm$1.3 & 3.0$\pm$0.3 & 0.66$\pm$0.25 & 264$\pm$~\,8 \\
138764 & IU Lib & B6 IV & 4.148$\pm$0.020 & 4.20$\pm$0.20 & 3.9$\pm$0.6 & 2.7$\pm$0.8 & 2.4$\pm$0.3 & 0.42$\pm$0.32 & 9$\pm$~\,5 \\
140873 & 25 Ser & B8 III, SB2 & 4.144$\pm$0.020 & 4.35$\pm$0.20 & 3.7$\pm$0.4 & 2.4$\pm$0.4 & 2.3$\pm$0.2 & 0.26$\pm$0.21 & 70$\pm$~\,2 \\
143309 & V350 Nor & B8/B9 Ib/II & 4.147$\pm$0.020 & 4.09$\pm$0.20 & 4.0$\pm$0.7 & 3.0$\pm$1.0 & 2.5$\pm$0.3 & 0.58$\pm$0.28 & --\,~~~ \\
160124 & V994 Sco & B3 IV-V, SB2 & 4.171$\pm$0.020 & 4.34$\pm$0.20 & 4.0$\pm$0.5 & 2.5$\pm$0.5 & 2.4$\pm$0.2 & 0.27$\pm$0.23 & 8$\pm$~\,4 \\
161783 & V539 Ara & B2 V, SB2 & 4.246$\pm$0.020 & 4.09$\pm$0.20 & 5.7$\pm$1.1 & 3.6$\pm$1.2 & 3.0$\pm$0.3 & 0.56$\pm$0.29 & --\,~~~ \\
169820 & BD+14 3533 & B9 V & 4.071$\pm$0.020 & 4.26$\pm$0.20 & 2.9$\pm$0.4 & 2.2$\pm$0.6 & 1.9$\pm$0.3 & 0.37$\pm$0.29 & --\,~~~ \\
177863 & V4198 Sgr & B8 V, SB1 & 4.127$\pm$0.020 & 4.14$\pm$0.20 & 3.7$\pm$0.6 & 2.8$\pm$0.9 & 2.3$\pm$0.3 & 0.51$\pm$0.30 & 63$\pm$~\,2 \\
181558 & V4199 Sgr & B5 III & 4.167$\pm$0.020 & 4.16$\pm$0.20 & 4.2$\pm$0.7 & 2.9$\pm$1.0 & 2.5$\pm$0.3 & 0.46$\pm$0.34 & 6$\pm$~\,4 \\
182255 & 3 Vul & B6 III, SB1 & 4.150$\pm$0.020 & 4.23$\pm$0.20 & 3.9$\pm$0.6 & 2.6$\pm$0.7 & 2.4$\pm$0.3 & 0.39$\pm$0.31 & 10$\pm$~\,6 \\
206540 & BD+10 4604 & B5 IV & 4.146$\pm$0.020 & 4.11$\pm$0.20 & 4.0$\pm$0.7 & 3.0$\pm$1.0 & 2.5$\pm$0.3 & 0.55$\pm$0.29 & 8$\pm$~\,5 \\
208057 & 16 Peg & B3 V, SB ? & 4.222$\pm$0.020 & 4.15$\pm$0.20 & 5.1$\pm$0.8 & 3.2$\pm$1.0 & 2.8$\pm$0.3 & 0.46$\pm$0.33 & 104$\pm$~\,6 \\
215573 & $\xi$ Oct & B6 IV & 4.145$\pm$0.020 & 4.09$\pm$0.20 & 4.0$\pm$0.7 & 3.0$\pm$1.0 & 2.5$\pm$0.3 & 0.58$\pm$0.28 & 5$\pm$~\,2 \\
\hline
\end{tabular}
\end{center}
\end{table*}
\begin{table*}
\caption{
The mean longitudinal field measurements for our sample
of $\beta$~Cephei and SPB stars observed
with FORS\,1 (ESO service programs
072.D-0377, 073.D-0464, 073.D-0466, and 075.D-0295).
The first two columns list the HD number
and the modified Julian date of mid-exposure.
The measured mean longitudinal magnetic field $\left<B_{\mathrm l}\right>$
is presented in column~3.
If there are several measurements for a single star,
we give the rms longitudinal magnetic field
and the reduced $\chi^2$ for all measurements in columns~4 and 5.
}
\label{table2}
\begin{center}
\begin{tabular}{rcrrrp{2mm}rcrrr}
\cline{1-5} \cline{7-11}
\multicolumn{1}{c}{\raisebox{2mm}{\rule{0mm}{2mm}}HD} &
\multicolumn{1}{c}{MJD} &
\multicolumn{1}{c}{$\left<B_{\mathrm l}\right>$} &
\multicolumn{1}{c}{$\overline{\left< B_l \right>}$} &
\multicolumn{1}{c}{$\chi^2/n$} &
 &
\multicolumn{1}{c}{HD} &
\multicolumn{1}{c}{MJD} &
\multicolumn{1}{c}{$\left<B_{\mathrm l}\right>$} &
\multicolumn{1}{c}{$\overline{\left< B_l \right>}$} &
\multicolumn{1}{c}{$\chi^2/n$} \\
\multicolumn{1}{c}{} &
\multicolumn{1}{c}{} &
\multicolumn{1}{c}{[G]} &
\multicolumn{1}{c}{[G]} &
\multicolumn{1}{c}{} &
 &
\multicolumn{1}{c}{} &
\multicolumn{1}{c}{} &
\multicolumn{1}{c}{[G]} &
\multicolumn{1}{c}{[G]} &
\multicolumn{1}{c}{} \\
\cline{1-5} \cline{7-11} 
3379 & 53244.402 & 272$\pm$ 57 & 182 & 13.5 & &	129929 & 53454.199 & -52$\pm$ 38 & 67 & 3.5 \\
 & 53245.214 & 231$\pm$ 47 & & & &	 & 53572.053 & -80$\pm$ 35 & & \\
 & 53629.305 & 5$\pm$ 22 & & & &	131058 & 53454.220 & -106$\pm$ 46 &   &   \\
 & 53630.195 & -67$\pm$ 25 & & & &	138764 & 52904.016 & 146$\pm$ 57 &   &   \\
24587 & 52971.071 & -120$\pm$ 68 & 72 & 1.5 & &	140873 & 53151.192 & 286$\pm$ 60 & 231 & 13.4 \\
 & 53574.415 & -16$\pm$ 36 & & & &	 & 53454.247 & -158$\pm$ 78 & & \\
 & 53630.250 & -32$\pm$ 29 & & & &	143309 & 53151.192 & -196$\pm$ 90 & 113 & 1.7 \\
26326 & 52909.389 & -119$\pm$ 80 & 72 & 1.3 & &	 & 53225.056 & 85$\pm$ 95 & & \\
 & 53218.357 & -24$\pm$ 99 & & & &	 & 53234.019 & -75$\pm$ 73 & & \\
 & 53630.370 & -27$\pm$ 22 & & & &	 & 53454.280 & -8$\pm$ 35 & & \\
28114 & 53638.390 & -58$\pm$ 21 &   &   & &	157056 & 53532.324 & -39$\pm$ 21 &   &   \\
29248 & 53629.322 & -41$\pm$ 28 & 33 & 1.6 & &	160124 & 53151.259 & 456$\pm$ 60 & 232 & 15.5 \\
 & 53630.347 & -22$\pm$ 21 & & & &	 & 53520.230 & -77$\pm$ 45 & & \\
34798 & 52999.055 & 56$\pm$ 82 & 125 & 3.6 & &	 & 53600.109 & 19$\pm$ 36 & & \\
 & 53218.406 & -205$\pm$ 98 & & & &	 & 53604.108 & -25$\pm$ 24 & & \\
 & 53638.366 & -41$\pm$ 17 & & & &	160578 & 53532.342 & -45$\pm$ 41 & 73 & 3.3 \\
44743 & 53475.029 & 24$\pm$ 70 & 35 & 1.2 & &	 & 53604.127 & 93$\pm$ 40 & & \\
 & 53629.343 & -44$\pm$ 29 & & & &	161783 & 53151.281 & 376$\pm$ 63 & 210 & 14.6 \\
45284 & 53252.365 & 245$\pm$ 63 &   &   & &	 & 53487.333 & -88$\pm$ 31 & & \\
46005 & 53259.399 & 2$\pm$ 79 &   &   & &	 & 53520.308 & -113$\pm$ 32 & & \\
46328 & 53475.046 & 280$\pm$ 44 & 306 & 47.1 & &	 & 53598.108 & -122$\pm$ 78 & & \\
 & 53506.971 & 330$\pm$ 45 & & & &	163472 & 53151.298 & 121$\pm$ 54 &   &   \\
53921 & 52999.137 & -294$\pm$ 63 & 185 & 25.3 & &	169820 & 53151.312 & 239$\pm$ 70 & 147 & 5.3 \\
 & 53475.100 & -71$\pm$ 87 & & & &	 & 53520.333 & -86$\pm$ 45 & & \\
 & 53630.401 & 151$\pm$ 29 & & & &	 & 53597.112 & -26$\pm$ 34 & & \\
 & 53631.408 & 151$\pm$ 21 & & & &	177863 & 53193.211 & -21$\pm$ 54 & 44 & 2.7 \\
74195 & 53002.127 & -277$\pm$108 & 200 & 5.0 & &	 & 53597.128 & -59$\pm$ 26 & & \\
 & 53138.972 & -310$\pm$ 98 & & & &	181558 & 53193.251 & -114$\pm$ 50 & 201 & 8.4 \\
 & 53143.972 & -145$\pm$ 70 & & & &	 & 53227.184 & 236$\pm$ 75 & & \\
 & 53454.107 & -65$\pm$ 39 & & & &	 & 53274.144 & -247$\pm$ 98 & & \\
 & 53455.080 & -44$\pm$ 37 & & & &	 & 53275.143 & -336$\pm$ 63 & & \\
74560 & 53002.141 & -199$\pm$ 61 & 146 & 5.6 & &	 & 53519.376 & 0$\pm$ 34 & & \\
 & 53143.986 & -53$\pm$ 66 & & & &	 & 53520.397 & -26$\pm$ 31 & & \\
85953 & 53002.152 & -131$\pm$ 42 & 103 & 6.1 & &	182255 & 53193.234 & -13$\pm$ 57 &   &   \\
 & 53152.970 & -151$\pm$ 49 & & & &	206540 & 53514.416 & -51$\pm$ 31 &   &   \\
 & 53454.137 & -52$\pm$ 23 & & & &	208057 & 53192.307 & 104$\pm$ 66 & 133 & 13.9 \\
 & 53455.109 & 10$\pm$ 21 & & & &	 & 53597.166 & -156$\pm$ 31 & & \\
92287 & 53008.352 & -10$\pm$ 57 & 37 & 2.8 & &	215573 & 52900.080 & 165$\pm$ 53 & 174 & 6.9 \\
 & 53454.152 & -52$\pm$ 22 & & & &	 & 53191.222 & 180$\pm$ 54 & & \\
111123 & 53455.155 & -14$\pm$ 38 & 18 & 0.2 & &	 & 53192.290 & 123$\pm$ 66 & & \\
 & 53475.154 & 22$\pm$ 51 & & & &	 & 53193.321 & -320$\pm$ 90 & & \\
123515 & 52824.093 & -59$\pm$ 50 & 43 & 0.9 & &	 & 53506.414 & 60$\pm$ 45 & & \\
 & 53454.179 & 17$\pm$ 27 & & & &	 & 53522.420 & -36$\pm$ 21 & & \\
\cline{1-5} \cline{7-11} 
\end{tabular}
\end{center}
\end{table*}
\begin{figure*}
\centering
\includegraphics[width=0.30\textwidth]{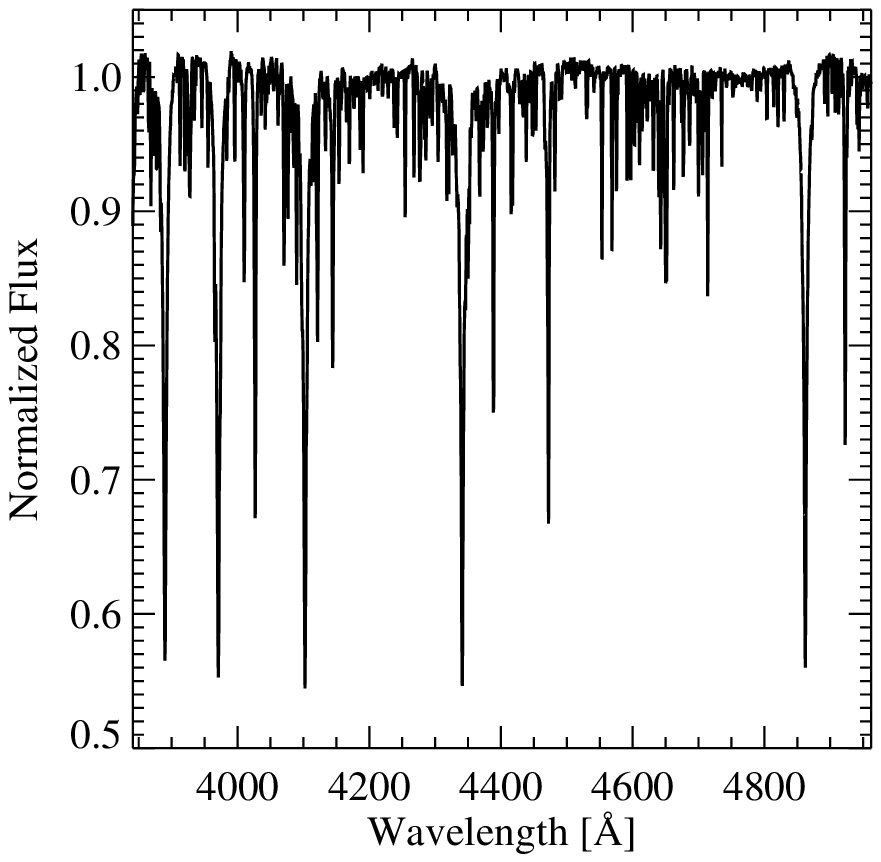}
\includegraphics[width=0.30\textwidth]{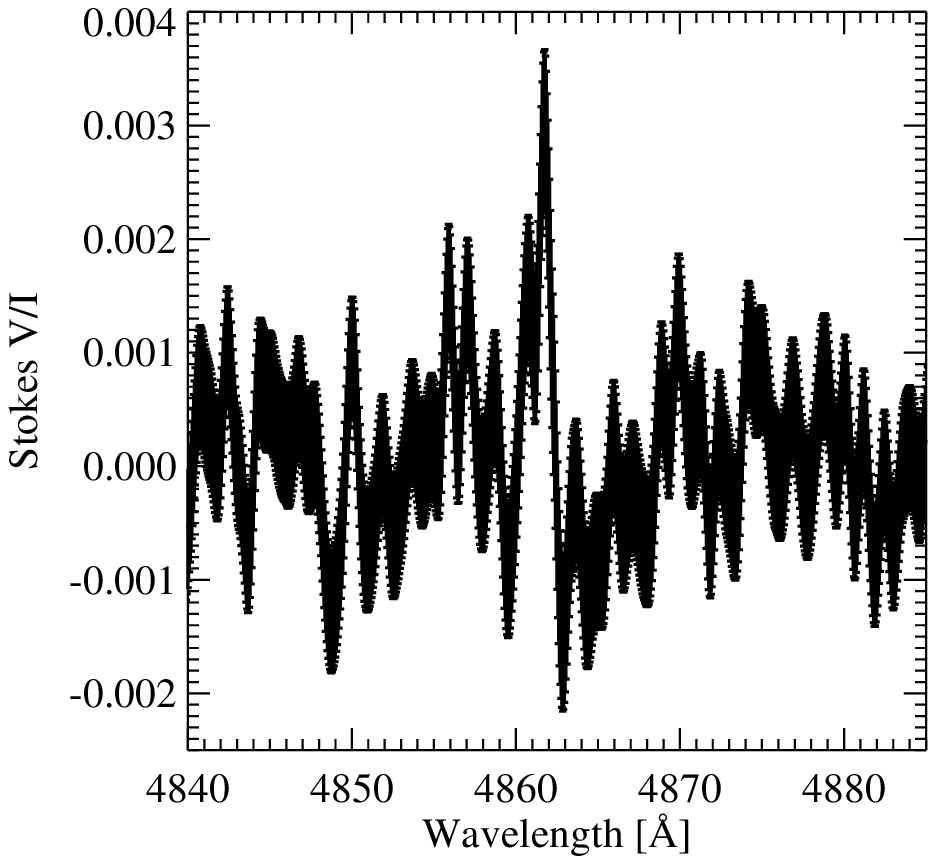}
\includegraphics[width=0.30\textwidth]{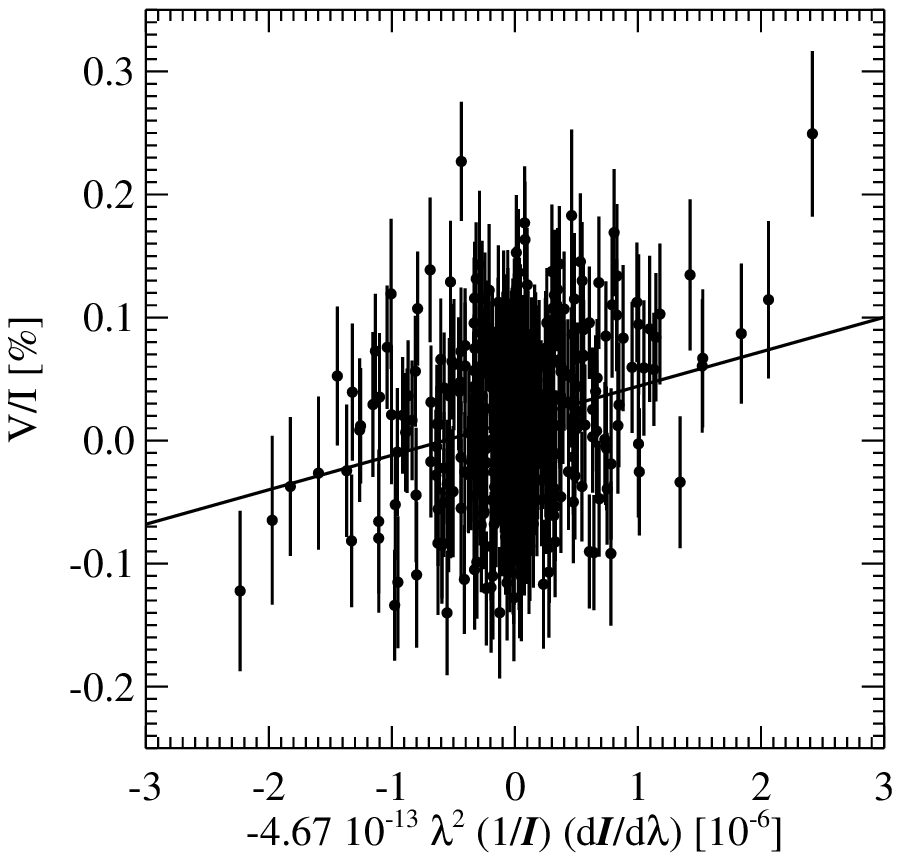}
\caption{Stokes I spectrum (left), Stokes V spectrum (centre) and regression
detection (right) for $\xi^1$\,CMa.  The thickness of the plotted line in the
Stokes V spectrum corresponds to the uncertainty of the measurement of
polarization determined from photon noise.  }
\label{fig1}  
\end{figure*}

\section{The sample of pulsating stars and observations}

Our sample includes eight $\beta$\,Cephei stars and 26 SPB stars
selected from \citet{DeCatproc2002} such as to fit the observational window and
having $V\le8$. Their fundamental parameters are listed in Table~\ref{table1}.
Observations in the \geneva{} photometric system are available for all
targets. The mean \geneva{} magnitudes were used to obtain the effective
temperature \logteff{} and the surface gravity \logg{} with the method described
in \citet{DeCatproc2002} (Columns~4 and 5 in Table~\ref{table1}).  The
\logteff{} and \logg{} values of HD\,44743, HD\,46005, and HD\,46328 are
inaccurate because an extrapolation outside the calibration grid was needed.
Other stellar parameters were derived from a grid of main-sequence models
calculated with the Code Li\'egeois d'\'Evolution Stellaire (version 18.2,
written by R.\ Scuflaire) described as ``grid 2'' in \citet{DeCat2006}.  The
mass $M$, the radius $R$, the luminosity \logl, and the age of the star
expressed as a fraction of its total main-sequence life $f$ are presented in
Columns~6 to 9 in Table~\ref{table1}.  For the most massive stars of our sample,
HD\,44743, HD\,46328, and HD\,111123, the upper value of 15\,$M_\odot$ of the
grid models is insufficient to fully cover the observed errorbox in \logteff{}
and \logg{}.  The stars HD\,45284 and HD\,46005 fall outside the main sequence.
The uncertain parameters are given in {\it italics} in Table~\ref{table1}.  In
addition, we point out that the physical parameters of multiple systems
(indicated as SB1 and SB2 in Column\,3 in Table~\ref{table1}), also have to be
treated with caution.

For most targets numerous high-resolution spectroscopic time series were
obtained in previous years
(\citealt{AertsDeCat2003,DeCat2002,Aerts2004a,Aerts2004b}).  To estimate the
projected rotational velocity \vsini{} (Column~10 in Table~\ref{table1}), an
average of all spectra has been used.  For slowly rotating $\beta$\,Cephei and
SPB stars, we selected several unblended absorption lines: the $\lambda$
4560\,\AA{} Si\,III-triplet and/or the $\lambda$ 4130\,\AA{} Si\,II-doublet.
For rapidly rotating stars, only the $\lambda$ 4481\,\AA{} Mg\,I-line has been
used.  We applied the method of least squares fitting with
rotationally broadened synthetic profiles using a Gaussian intrinsic width but
without taking into account pulsational broadening.

The spectropolarimetric observations were carried out from 2003 to 2005 at the
European Southern Observatory with FORS\,1 (FOcal Reducer low dispersion
Spectrograph) mounted on the 8-m Melipal telescope of the VLT.  This multi-mode
instrument is equipped with polarization analyzing optics comprising
super-achromatic half-wave and quarter-wave phase retarder plates, and a
Wollaston prism with a beam divergence of 22$\arcsec$ in standard resolution
mode.  In 2003 and 2004, we used the GRISM\,600B in the wavelength range
3480--5890\,\AA{} to cover all hydrogen Balmer lines from H$\beta$ to the Balmer
jump.  The highest spectral resolution of the FORS\,1 spectra achieved with this
setting is $R\sim2000$.  As of April 2005, we used the GRISM\,1200g to cover the
H Balmer lines from H$\beta$ to H$8$, and the narrowest available slit width of
0$\farcs$4 to obtain a spectral resolving power of $R\sim4000$.  Usually, we
took four to eight continuous series of two exposures for each sample star using
a standard readout mode with high gain (A,1$\times$1,high) and with the retarder
waveplate oriented at two different angles, +45$^\circ$ and $-$45$^\circ$.  For
the observations carried out in 2005, we used a non-standard readout mode with
low gain (A,1$\times$1,low), which provided a broader dynamic range and
increased the signal-to-noise (S/N) ratio of individual spectra by a factor of
$\approx$2.  All sample stars are bright and, since the errors of the
measurements of the polarization with FORS\,1 are determined by photon counting
statistics, a S/N ratio of a few thousand was reached within $\sim$30\,min.
More details on the observing technique with FORS 1 can be found elsewhere
(e.g., \citealt{Bagnulo2002,Hubrig2004}).  Determination of the longitudinal
magnetic field using the FORS\,1 spectra is achieved by measuring the circular
polarization of opposite sign induced in the wings of hydrogen Balmer lines by
the Zeeman effect.  The mean longitudinal magnetic field is the average over the
stellar hemisphere visible at the time of observation of the component of the
field parallel to the line of sight, weighted by the local emergent spectral
line intensity.  It is diagnosed from the slope of a linear regression of $V/I$
versus the quantity $-g_{\rm eff} \Delta\lambda_z \lambda^2 \frac{1}{I}
\frac{{\mathrm d}I}{{\mathrm d}\lambda} \left<B_l\right > + V_0/I_0$, where $V$
is the Stokes parameter which measures the circular polarization, $I$ is the
intensity observed in unpolarized light, $g_{\rm eff}$ is the effective Land\'e
factor, $\lambda$ is the wavelength, ${{\rm d}I/{\rm d}\lambda}$ is the
derivative of Stokes $I$, and $\left<B_l\right>$ is the mean longitudinal field.
Our experience from a study of a large sample of magnetic and non-magnetic Ap
and Bp stars revealed that this regression technique is very robust and that
detections with $B_l > 3\,\sigma$ result only for stars possessing magnetic
fields.

\section{Analysis and results}

The mean longitudinal magnetic field $\left<B_l\right>$ of the targets is
listed in Table~\ref{table2}.  The rms longitudinal field is computed from all
$n$ measurements according to:
\begin{equation}
\overline{\left< B_l \right>} = \left( \frac{1}{n} \sum^{n}_{i=1} \left< B_l \right> ^2_i \right)^{1/2}.
\label{eqn1}
\end{equation}
Further we give the reduced $\chi^2$ for 
these measurements in Column~5, which is a statistical discriminant
useful to assess the presence of a magnetic field
\citep{Bohlender1993}, defined as:
\begin{equation}
\chi^2/n = \frac{1}{n} \sum_{i=1}^n \left( \frac{\left< B_l \right>_i}{\sigma_i} \right)^2.
\end{equation}
A longitudinal magnetic field at a level larger than 3\,$\sigma$ has been
diagnosed for 13 SPB stars: HD\,3379, HD\,45284, HD\,53921, HD\,74195,
HD\,74560, HD\,85953, HD\,140873, HD\,160124, HD\,161783, HD\,169820,
HD\,181558, HD\,208057, and HD\,215573. The star HD\,208057 has
a rather high \vsini{} value and was classified as a Be star by
\citet{MerrillBurwell1943} due to the detection of double emission in H$\alpha$.
Although the emissison was not confirmed in later observations, it cannot be
ruled out that this star is a Be star. Given the still uncertain
status of the magnetic field discovered by \citet{Neiner2003c} in $\omega$\,Ori,
HD\,208057 is possibly the first Be star with a magnetic field detected at
3\,$\sigma$ level.

Most stars with multiple measurements have $\chi^2/n\ge5.0$. The individual
measurements show the variability of their magnetic field. The time scales of
these variations are uncertain due to too few measurements being obtained for
each star.  Unfortunately, also the rotational variability of the magnetic field
cannot be proven as the rotational periods are not known for any of these
targets.  For the SPB star HD\,181558 with the largest number of magnetic field
measurements (six in all) we searched for a correlation between the changes of
the magnetic field with the pulsation period, but no clear correlation could be
detected.

The magnetic field of $\xi^1$\,CMa is detected at more than 6\,$\sigma$ level.
Both measurements carried out on two nights separated by about one month are of
the order of 300\,G. In Fig.\,\ref{fig1} we present Stokes I and V spectra and
the regression detection.  Spectroscopic and photometric
observations revealed that  $\xi^1$\,CMa 
pulsates monoperiodically and non-linearly in a radial mode
with a period of 0.209574 days, with a velocity amplitude of 
33\,km\,s$^{-1}$ \citep{Saesen2006}.
\citet{Morel2006} performed an abundance analysis of nine selected
$\beta$\,Cephei stars and discovered $\xi^1$\,CMa to be nitrogen enriched, as
well as the three $\beta\,$Cephei stars $\delta$\,Cet, V2052\,Oph, and
$\beta$\,Cep.  It is remarkable that two of them, V2052\,Oph and $\beta$\,Cep,
also have a detected longitudinal magnetic field of the order of $\sim$100\,G
\citep{Neiner2003b,Henrichs2000}.  $\xi^1$\,CMa shows the largest mean
longitudinal field by a factor of three among these three $\beta$\,Cephei stars.
Unfortunately, no search for a magnetic field has been attempted yet for
$\delta$ Cet.  These four stars have another common property: they are either
radial pulsators ($\xi^1$\,CMa) or their multiperiodic pulsations are dominated
by a radial mode ($\delta$\,Cet, $\beta$\,Cep, and V2052\,Oph).  The presence of
a magnetic field in these stars might play an important role to explain these
physical characteristics.
\begin{figure}
\centering
\includegraphics[height=0.44\textwidth,angle=270]{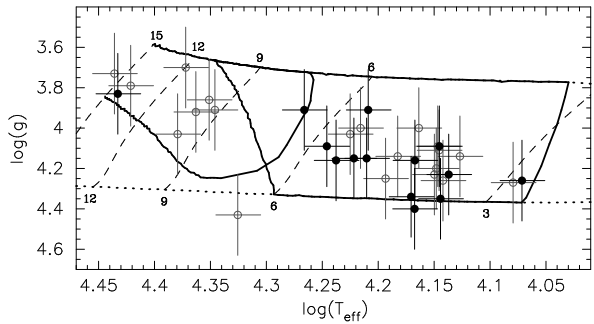}
\caption{
The position of the targets in the H-R diagram.
The full lines represent boundaries of theoretical instability strips
for modes with frequency between 0.2 and 30\,$\rm{d^{-1}}$
and $\ell \le 3$, computed for main sequence models with
$2\,M_{\odot}\le{}M\le{}15\,M_{\odot}$
of ``grid 2'' in \citet{DeCat2006}.
The lower and upper dotted lines show the ZAMS and TAMS, respectively.
The dashed lines denote evolution tracks for stars with $M$= 15, 12, 9, 6,
and 3\,$M_{\odot}$. Filled circles correspond to the stars with detected
magnetic fields.}  
\label{fig2}  
\end{figure}

The position of the studied SPB and $\beta$\,Cephei stars in the
H-\linebreak[0]R diagram is shown in Fig.\,\ref{fig2}.  No clear picture emerges
concerning the evolutionary stage of stars with detected magnetic fields.  We
have to note, however, that the uncertainties of the age of hot stars
(Table~\ref{table1}, Column~9) are very large.  Furthermore, the whole sample
under study contains only 14 stars with detected magnetic fields, so there is a
need for more magnetic field measurements to have a good statistics.  No hints
of relations between the magnetic field strength and other stellar parameters
were found.  Despite our discovery, the knowledge of the magnetic fields in
pulsating B stars remains very incomplete.  Magnetic fields are known to play an
important role in the theoretical interpretation of the pulsations in cool
rapidly oscillating Ap stars.  It is difficult to explain why non-pulsating
chemically peculiar hot Bp stars and pulsating stars co-exist in the SPB and
$\beta$\,Cephei instability strips. It is especially intriguing that the
magnetic fields of hot Bp stars either do not show any detectable variations or
vary with periods close to one day, which is of the order of the pulsation
period range of SPB stars \citep{Bohlender1987,Matthews1991}.  The presented
magnetic field measurements in 13 SPB stars and in the $\beta$\,Cephei star
$\xi^1$\,CMa demonstrate that longitudinal magnetic fields in these stars are
rather weak in comparison to the kG fields detected in magnetic Bp stars.

Although the magnetic field determination method based on circular polarized
FORS\,1 spectra of hydrogen Balmer lines shows the excellent potential of
FORS\,1 for the detection of weak fields, it would be particularly important to
measure the fields with higher resolution spectropolarimeters, not only using
hydrogen lines but also lines of other chemical elements to be able to study the
field configuration at high confidence level.

The role of the detected magnetic fields in the modelling of the oscillations
of B-type stars remains to be studied.  The magneto-acoustic coupling in
pulsating B stars will be far less important than for the roAp stars. The effect
on the p modes of the $\beta$\,Cephei stars is expected to be small
\citep{HasanChristensenDalsgaard1992} but the g modes of the SPB stars may be
slightly affected, provided that the internal field strength is a factor 1000
larger than the detected value \citep{Hasan2005}.  In any case, magnetic
breaking and angular momentum transport along the field lines now offer a
natural explanation of the slow rotation of our target stars.


\label{lastpage}

\end{document}